\documentclass{PoS}

\usepackage[utf8]{inputenc}
\usepackage{amsmath}
\usepackage{dcolumn}

\title{Nucleon form factors with light Wilson quarks}

\ShortTitle{Nucleon form factors with light Wilson quarks}

\author{\speaker{Jeremy~Green},$^a$\footnote{Current affiliation:
    \emph{Institut für Kernphysik, Johannes Gutenberg-Universität
      Mainz, D-55099 Mainz, Germany}}\ \ Michael~Engelhardt,$^b$
  Stefan~Krieg,$^{cd}$ Stefan~Meinel,$^a$ John~Negele,$^a$
  Andrew~Pochinsky$^a$ and Sergey~Syritsyn$^e$\\
  \llap{$^a$}Center for Theoretical Physics, Massachusetts Institute of Technology,\\
  Cambridge, Massachusetts 02139, USA\\
  \llap{$^b$}Department of Physics, New Mexico State University,
  Las Cruces, New Mexico 88003, USA\\
  \llap{$^c$}Bergische Universität Wuppertal,
  D-42119 Wuppertal, Germany\\
  \llap{$^d$}IAS, Jülich Supercomputing Centre, Forschungszentrum Jülich,
  D-52425 Jülich, Germany\\
  \llap{$^e$}Lawrence Berkeley National Laboratory,
  Berkeley, California 94720, USA\\
  E-mail: \email{green@kph.uni-mainz.de}, \email{engel@nmsu.edu},
  \email{s.krieg@fz-juelich.de}, \email{smeinel@mit.edu}, 
  \email{negele@mit.edu}, \email{avp@mit.edu}, \email{ssyritsyn@lbl.gov}}

\abstract{We present nucleon observables --- primarily isovector
  vector form factors --- from calculations using $2+1$~flavors of
  Wilson quarks. One ensemble is used for a dedicated high-precision
  study of excited-state effects using five source-sink separations
  between 0.7 and 1.6~fm. We also present results from a larger set of
  calculations that include an ensemble with pion mass 149~MeV and box
  size 5.6~fm, which nearly eliminates the uncertainty associated with
  extrapolation to the physical pion mass. The results show agreement
  with experiment for the vector form factors, which occurs only when
  excited-state contributions are reduced. Finally, we show results
  from a subset of ensembles that have pion mass 254~MeV with varying
  temporal and spatial box sizes, which we use for a controlled study
  of finite-volume effects and a test of the ``$m_\pi L=4$'' rule of
  thumb.}

\FullConference{31st International Symposium on Lattice Field Theory LATTICE 2013\\
		 July 29 – August 3, 2013\\
		 Mainz, Germany}

\begin{document}

\section{Introduction}

The isovector Dirac and Pauli form factors are defined via nucleon
matrix elements of the vector current:
\begin{equation}
  \langle N(p')|\bar\psi \gamma^\mu\tau^a \psi|N(p)\rangle = \bar u(p')\left( \gamma^\mu F_1^v(Q^2) + \frac{i\sigma^{\mu\nu}q_\nu}{2m}F_2^v(Q^2) \right)\tau^a\bar u(p),
\end{equation}
where $\psi^T=(u,d)$, $q=p'-p$, and $Q^2=-q^2$. For connecting with
experiment, this means that
\begin{equation}
  F_{1,2}^v = F_{1,2}^p - F_{1,2}^n,
\end{equation}
where $F_{1,2}^{p,n}$ are form factors of the electromagnetic current
in a proton and in a neutron. Near zero momentum transfer, these
contain the isovector Dirac and Pauli radii and anomalous magnetic
moment:
\begin{align}
  F_1^v(Q^2) &= 1 - \frac{1}{6}(r_1^2)^vQ^2 + \mathcal{O}(Q^4)\\
  F_2^v(Q^2) &= \kappa^v\left(1 - \frac{1}{6}(r_2^2)^vQ^2 + \mathcal{O}(Q^4)\right).
\end{align}

In recent years there has been increased attention paid to the problem
of excited-state contamination in lattice QCD calculations of nucleon
structure observables. These arise when the Euclidean time separations
between the nucleon source and the vector current, and between the
vector current and the nucleon sink, are too small to effectively
filter out other states with the same quantum numbers as the
ground-state nucleon. In Sec.~\ref{sec:hiprec}, we report results from
a high-precision study of excited-state effects using calculations
with multiple source-sink separations, which allow for testing
different methods for computing matrix elements.

We also report results from a separate set of calculations using
eleven lattice ensembles with a range of pion masses and lattice
volumes. We show in Sec.~\ref{sec:BMW} that good agreement with
experiment for the isovector vector form factors is achieved only when
excited states are under reasonable control and the pion mass is near
the physical point. In Sec.~\ref{sec:corners}, we describe the results
of a controlled study of finite-volume effects using four ensembles
with $m_\pi=254$~MeV.

\section{\label{sec:hiprec}High-precision study of excited-state effects}

For studying excited-state effects, we use a single ensemble generated
by USQCD with $2+1$ flavors of Wilson-clover fermions coupled to gauge
fields smeared with one level of stout smearing, as was used for
$N_f=3$ ensembles in Ref.~\cite{Beane:2012vq}. This
ensemble has lattice volume $32^3\times 96$, lattice spacing $a\approx
0.11$~fm, and pion mass $m_\pi\approx 317$~MeV. On it, we compute
nucleon observables using five source-sink separations
$T/a\in\{6,8,10,12,14\}$, covering a range between 0.7 and
1.6~fm. High precision is achieved by performing 24672 measurements on
1028 gauge configurations.

\begin{figure}
  \centering
\begin{minipage}{0.746\textwidth}
  \includegraphics[width=\columnwidth]{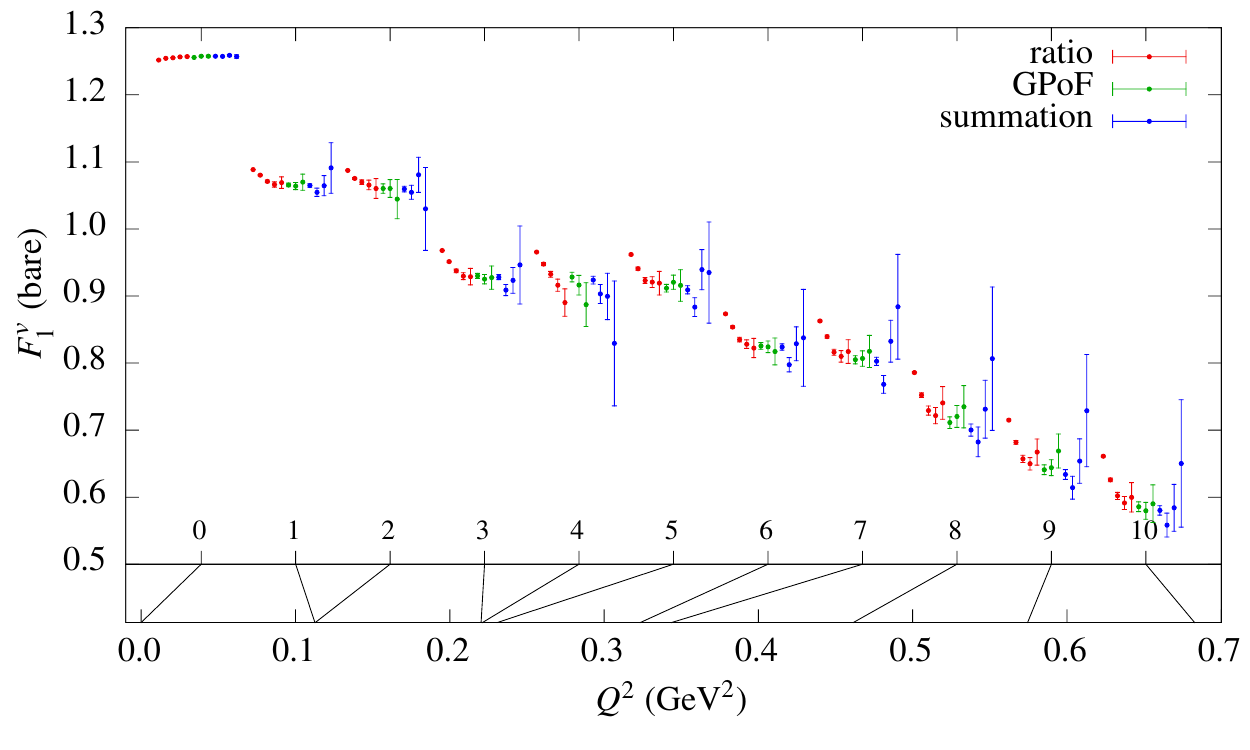}\\
  \includegraphics[width=\columnwidth]{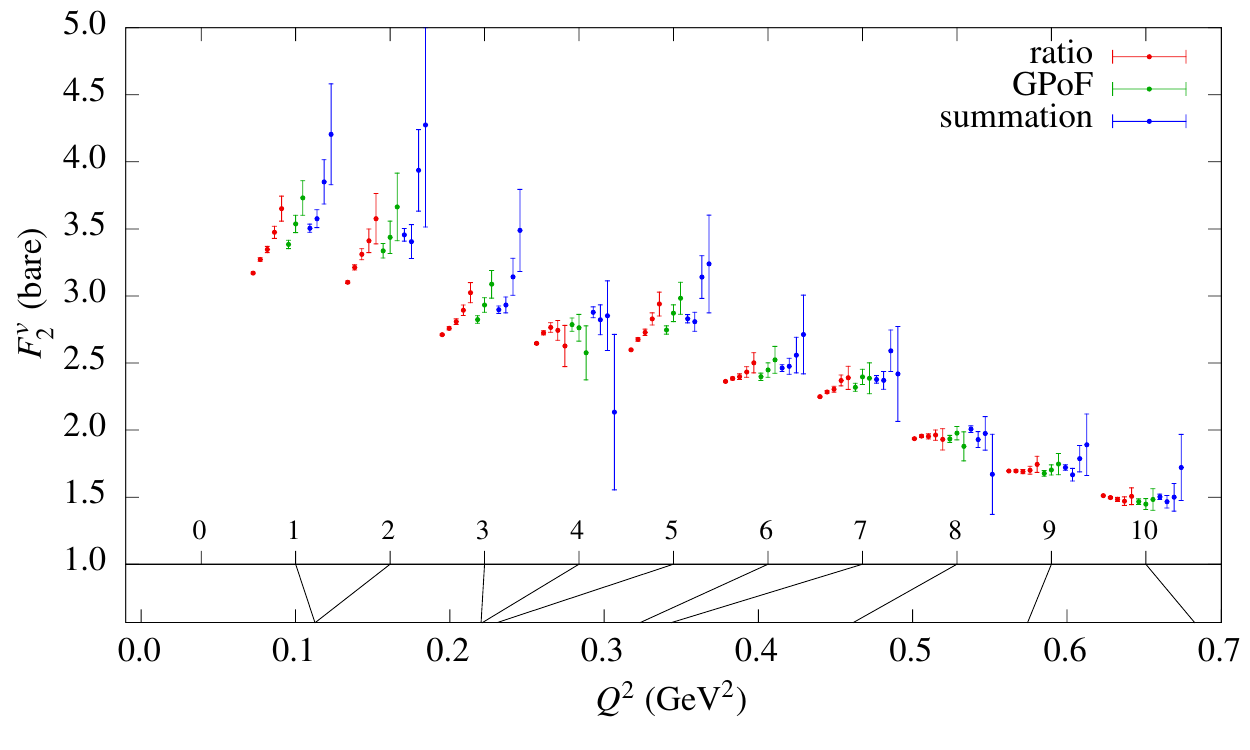}
\end{minipage}
{\small
\begin{tabular}{c|D{|}{|\:|}{7.6}}
\#& \langle \vec n'|\vec n\rangle\\\hline
0 & \langle  0,0,0|0,0,0\rangle\\
1 & \langle  0,0,0|1,0,0\rangle\\
2 & \langle -1,0,0|{-1},1,0\rangle\\
3 & \langle  0,0,0|1,1,0\rangle\\
4 & \langle -1,0,0|{-1},1,1\rangle\\
5 & \langle -1,0,0|0,1,0\rangle\\
6 & \langle  0,0,0|1,1,1\rangle\\
7 & \langle -1,0,0|0,1,1\rangle\\
8 & \langle -1,0,0|1,0,0\rangle\\
9 & \langle -1,0,0|1,1,0\rangle\\
10& \langle -1,0,0|1,1,1\rangle
\end{tabular}
}
  \caption{Isovector Dirac and Pauli form factors $F_1^v(Q^2)$ and
    $F_2^v(Q^2)$ on the USQCD ensemble, computed using the ratio,
    GPoF, and summation methods. Each method is shown with multiple
    points at each $Q^2$, corresponding to different source-sink
    separations increasing from left to right. The table lists
    representative source and sink momenta ($\vec
    p=\frac{2\pi}{L_s}\vec n$ and $\vec p'=\frac{2\pi}{L_s}\vec n'$,
    respectively) for each momentum transfer $Q^2$.}
  \label{fig:hiprec}
\end{figure}

We use three different methods for computing matrix elements from
two-point and three-point correlators:
\begin{enumerate}
\item The traditional \emph{ratio-plateau} method. For each
  source-sink separation, averaging over a fixed number of points at
  the center of each plateau yields a result with excited-state
  contributions asymptotically falling off as $e^{-\Delta
    E^\text{min}_{10}T/2}$, where $\Delta
  E^\text{min}_{10}=\min\{\Delta E_{10}(\vec p),\Delta E_{10}(\vec
  p')\}$ and the latter are the energy gaps between the ground state
  and the first excited state at the source and at the sink.
\item The \emph{generalized pencil-of-function}, or GPoF,
  method~\cite{Aubin:2010jc}. This is based on the recognition that
  the time-displaced nucleon interpolating operator $N^\tau(t)\equiv
  N(t+\tau)$ is linearly independent from $N(t)$. Therefore, by
  combining three-point correlators with three equally-spaced
  source-sink separations, we can use the variational
  method~\cite{Luscher:1990ck,Blossier:2009kd} to find linear
  combinations of $N$ and $N^\tau$ that asymptotically eliminate the
  lowest-lying excited state. Applying the ratio-plateau method to the
  result yields the ground-state matrix element with excited-state
  contributions that behave like $e^{-\Delta E^\text{min}_{20}T/2}$.
\item The \emph{summation} method involves using the ratio method,
  summing over the operator-insertion time, and finding the dependence
  of the resulting sums on the source-sink separation. This requires
  at least two source-sink separations, but asymptotically reduces the
  excited-state errors to $Te^{-\Delta
    E^\text{min}_{10}T}$~\cite{Capitani:2010sg,Bulava:2010ej}.
\end{enumerate}
With the data collected on this ensemble, the dependence on
source-sink separation of each of these methods can be probed: the
five source-sink separations yield five different ratio-method
results; using GPoF with $\tau=2a$ makes use of three adjacent
source-sink separations, which can be the lowest, middle, or largest
three to yield three different results; and taking the difference
between sums at adjacent source-sink separations yields four different
summation-method results. The resulting Dirac and Pauli form factors
computed using these methods are shown in Fig.~\ref{fig:hiprec}.

The effect of excited-state contamination can be clearly seen in the
ratio-method data at low source-sink separations. The value of
$F_1(Q^2)$ tends to decrease as excited states are removed by
increasing the source-sink separation, and this effect grows stronger
at larger $Q^2$. The result is that the value of the Dirac radius,
given by the slope of $F_1(Q^2)$ at $Q^2=0$, tends to increase as
excited-state effects are removed.

At low $Q^2$, the value of $F_2(Q^2)$ shows a strong increase with the
source-sink separation; this effect decreases at larger $Q^2$, and
between $0.5$ and $0.7\text{ GeV}^2$ it appears to change sign. This
means that the slope and the intercept, extrapolated to $Q^2=0$, will
both grow as excited-state effects are removed, causing the computed
values of the magnetic moment and Pauli radius to increase.

For the Dirac form factor, the ratio method appears to stabilize at a
plateau by the third or fourth source-sink separation, i.e., around
1.1 to 1.3~fm. The ratio data for the Pauli factor at low $Q^2$
generally don't appear to be approaching a plateau, although it is
possible that the apparent accelerating growth with the source-sink
separation may simply be a random fluctuation of correlated data;
comparing momentum transfer \#4 with the adjacent two suggests that
this may be the case.

The GPoF method produces results that are quite similar
to what is obtained from the ratio method using the largest of the
three source-sink separations used for GPoF, although with slightly
larger errors.

The power of the summation method for eliminating excited-state
contamination can be seen by considering the smallest two source-sink
separations. Using the ratio method, this produces the first two ratio
points in the plots, which suffer from significant excited-state
effects. However, the same data can be combined to produce the first
set of summation points, which are generally compatible with the
plateau ultimately obtained using the ratio method (for $F_1$) or with
the fourth ratio-method point (for $F_2$). On the other hand, the
summation method produces rather large statistical errors, so it is
still not clear as to whether the summation method is generally
superior to simply using the ratio method at larger source-sink
separations.

\section{\label{sec:BMW}Isovector vector form factors near the physical pion mass}

Our larger set of calculations uses the action developed by the BMW
collaboration, with $2+1$~flavors of Wilson-clover fermions coupled to
gauge fields smeared with two levels of HEX
smearing~\cite{Durr:2010aw}. We use mostly ``coarse'' ensembles with
$a=0.116$~fm and pion masses ranging from the near-physical 149~MeV to
356~MeV, with spatial box sizes that mostly satisfy $m_\pi L_s\gtrsim
4$.  On each ensemble, we compute three-point correlators using three
source-sink separations that are close to the middle three used for
the high-precision study in the previous section. Because the data are
noisier than in the previous section, we use the summation method in
just one way, fitting a line to the sums for the three source-sink
separations.

With these ensembles, we performed dipole fits in the range $0\leq
Q^2<0.5\text{ GeV}^2$ to determine the Dirac and Pauli radii and the
anomalous magnetic moment. Using the summation method and
extrapolating these quantities to the physical pion mass produced
agreement with experiment~\cite{Green:2012ud}, and their values on the
ensemble with $m_\pi=149$~MeV and $L_s=5.6$~fm were also close to the
experimental values.
\begin{figure}
  \centering
  \includegraphics[width=0.495\columnwidth]{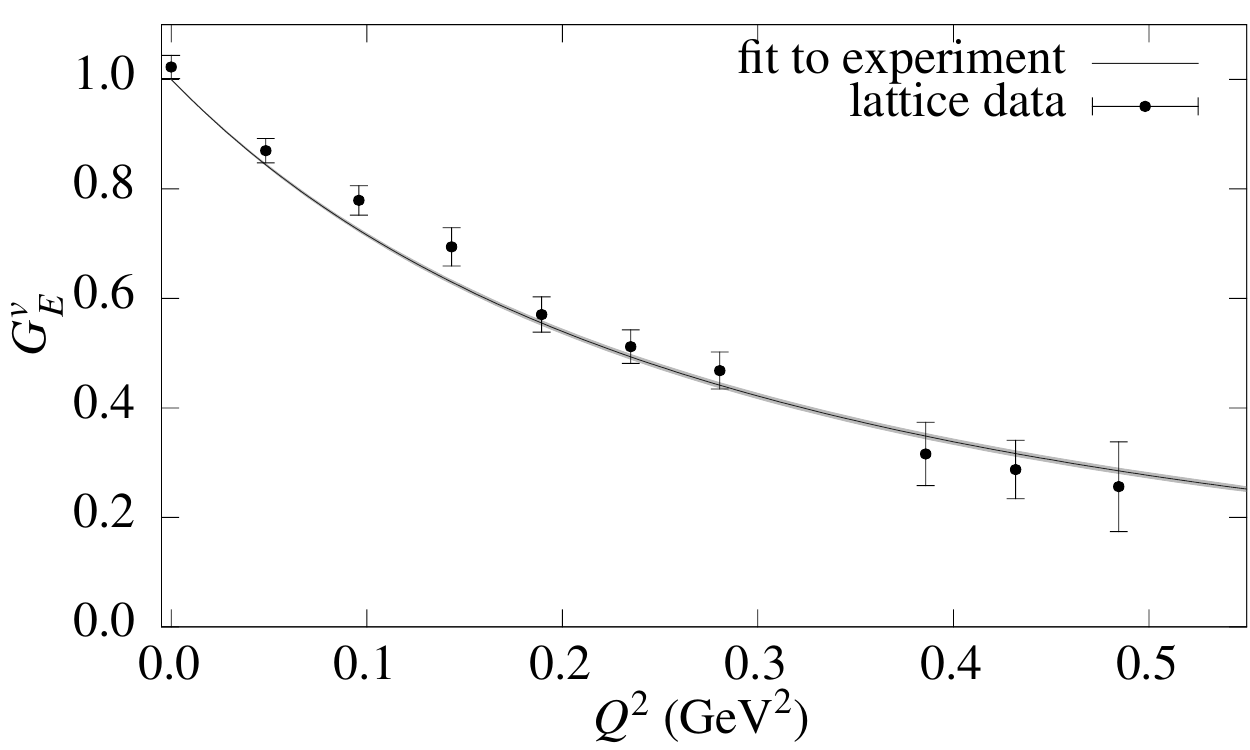}
  \includegraphics[width=0.495\columnwidth]{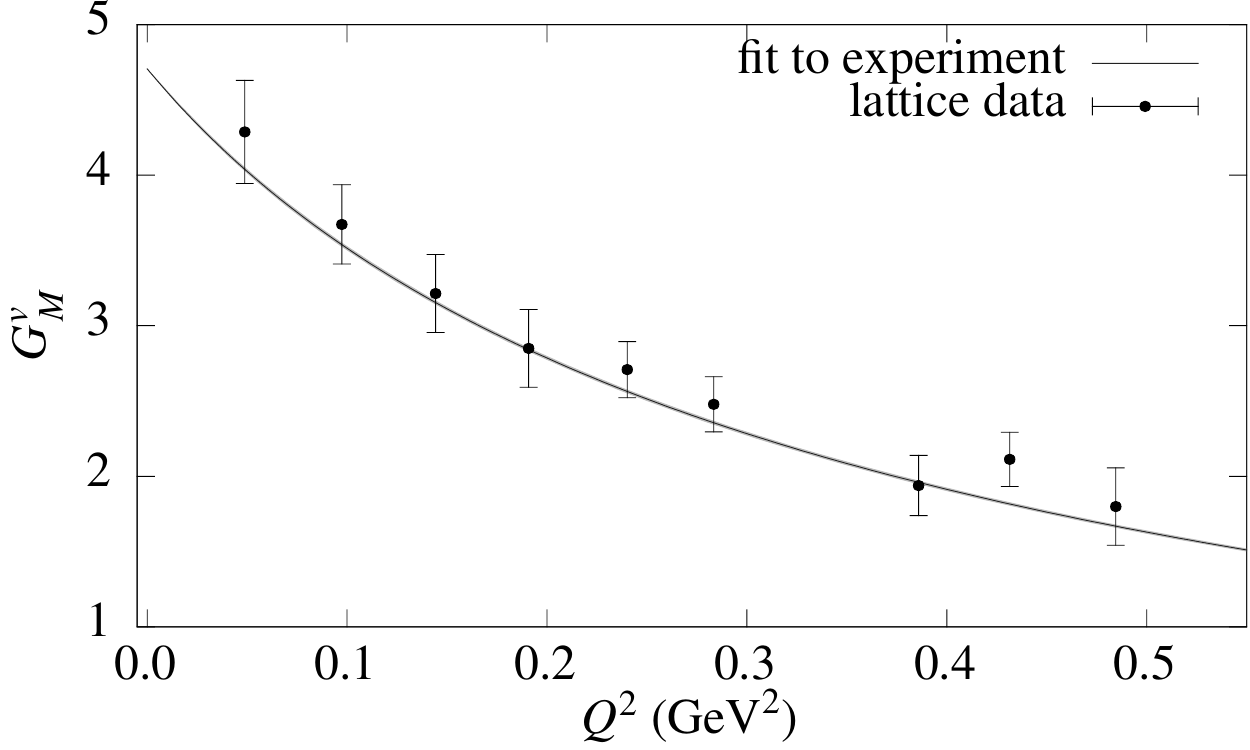}
  \caption{Isovector electric and magnetic form factors, computed on
    the $m_\pi=149$~MeV ensemble using the summation method (data at
    nearby $Q^2$ are binned for clarity). The curves are from the fit
    to experiment in Ref.~\cite{Alberico:2008sz}.}
  \label{fig:sachs}
\end{figure}

To avoid the complications involved in determining behavior at
$Q^2=0$, it is interesting to compare the form factors themselves with
experiment. To that end, we plot the Sachs form factors,
\begin{align}
  G_E^v(Q^2) &= F_1^v(Q^2) - \frac{Q^2}{2m_N}F_2^v(Q^2) \\
  G_M^v(Q^2) &= F_1^v(Q^2) + F_2^v(Q^2),
\end{align}
from the $m_\pi=149$~MeV ensemble together with experimental curves
that include experimental uncertainties~\cite{Alberico:2008sz}, in
Fig.~\ref{fig:sachs}. The lattice data are consistent with experiment
($p=0.64$ for $G_E$ and $p=0.81$ for $G_M$), but this is only achieved
with both reasonable control over excited states and a near-physical
pion mass. If either of these conditions is not satisfied, then the
agreement fails; for example, using the ratio method with $T=1.16$~fm
or using an ensemble with $m_\pi=254$~MeV yields $p<10^{-3}$.

\section{\label{sec:corners}Controlled study of finite-volume effects}

A particular subset of the ensembles used for the calculations
described in the previous section allows for a controlled study of
finite-volume effects. This consists of four ensembles with
$m_\pi=254$~MeV that differ only in their space and time extents: with
$a=0.116$~fm, these are $24^3\times 24$, $24^3\times 48$, $32^3\times
24$, and $32^3\times 48$. We look for the volume dependence of a
given observable by fitting
\begin{equation}\label{eq:corners_fit}
A + Be^{-m_\pi L_s} + Ce^{-m_\pi L_t}
\end{equation}
to the four data points. We then use this for interpolating to test the
``$m_\pi L=4$'' rule of thumb and determine the relative error caused
by $m_\pi L_s=4$ and $m_\pi L_t=4$ via $e^{-4}B/A$ and $e^{-4}C/A$,
respectively.

Using the summation-method results, we find that finite-volume errors
are consistent with zero for $(r_1^2)^v$, $(r_2^2)^v$, and $\kappa^v$,
although the statistical uncertainties are moderately large: for
$e^{-4}B/A$, they are roughly 0.1 for $(r_1^2)^v$ and $\kappa^v$, and
0.2 for $(r_2^2)^v$; the corresponding statistical uncertainties for
$e^{-4}C/A$ are slightly smaller.

Reduced statistical uncertainties can be obtained by using the ratio
method, although there is the possibility that volume dependence could
be caused by volume-dependent excited-state effects (such as those
from multiparticle states), rather than a finite-volume effect in the
ground state. For $(r_1^2)^v$, the ratio method at $T=0.93$~fm still
shows no sign of finite-volume effects, and the statistical
uncertainty at $m_\pi L=4$ is reduced to 0.02. On the other hand, the
same method indicates that $\kappa^v$ and $(r_2^2)^v$ suffer from a
$-5\%$ shift at $m_\pi L_s=4$ and a similar shift in the opposite
direction at $m_\pi L_t=4$.

\begin{figure}
  \centering
  \includegraphics[width=0.495\columnwidth]{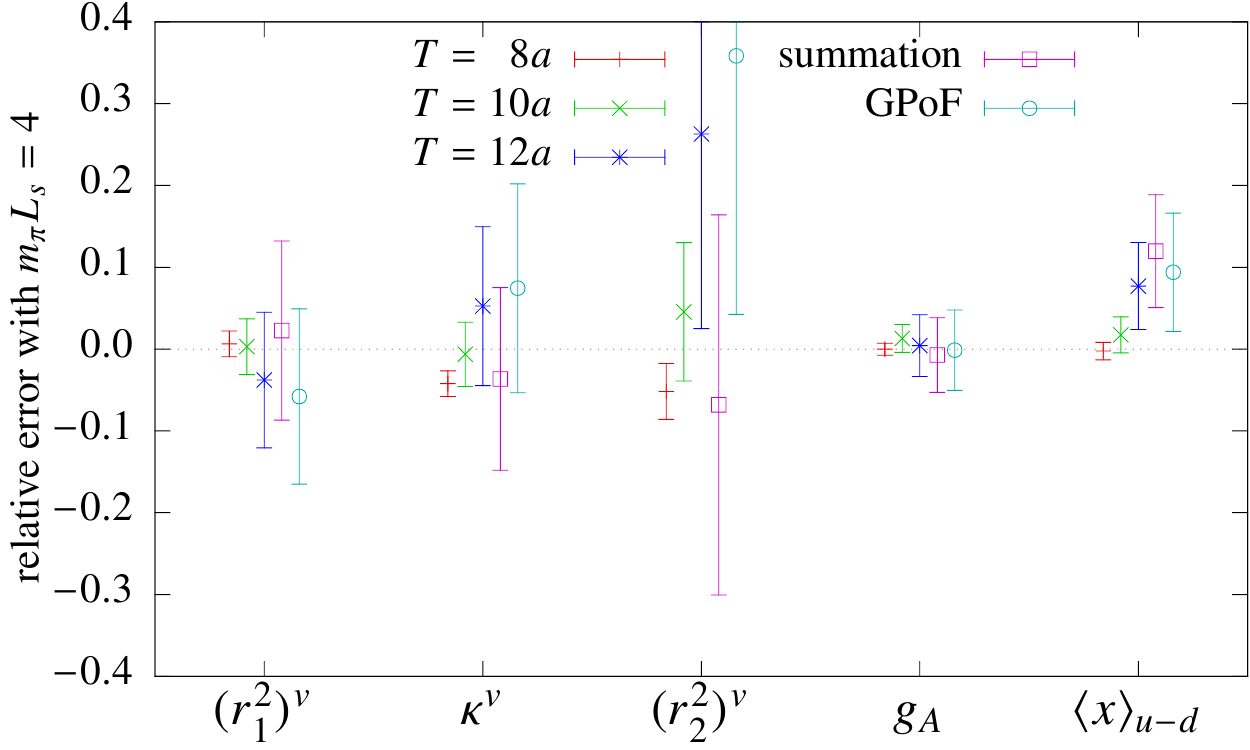}
  \includegraphics[width=0.495\columnwidth]{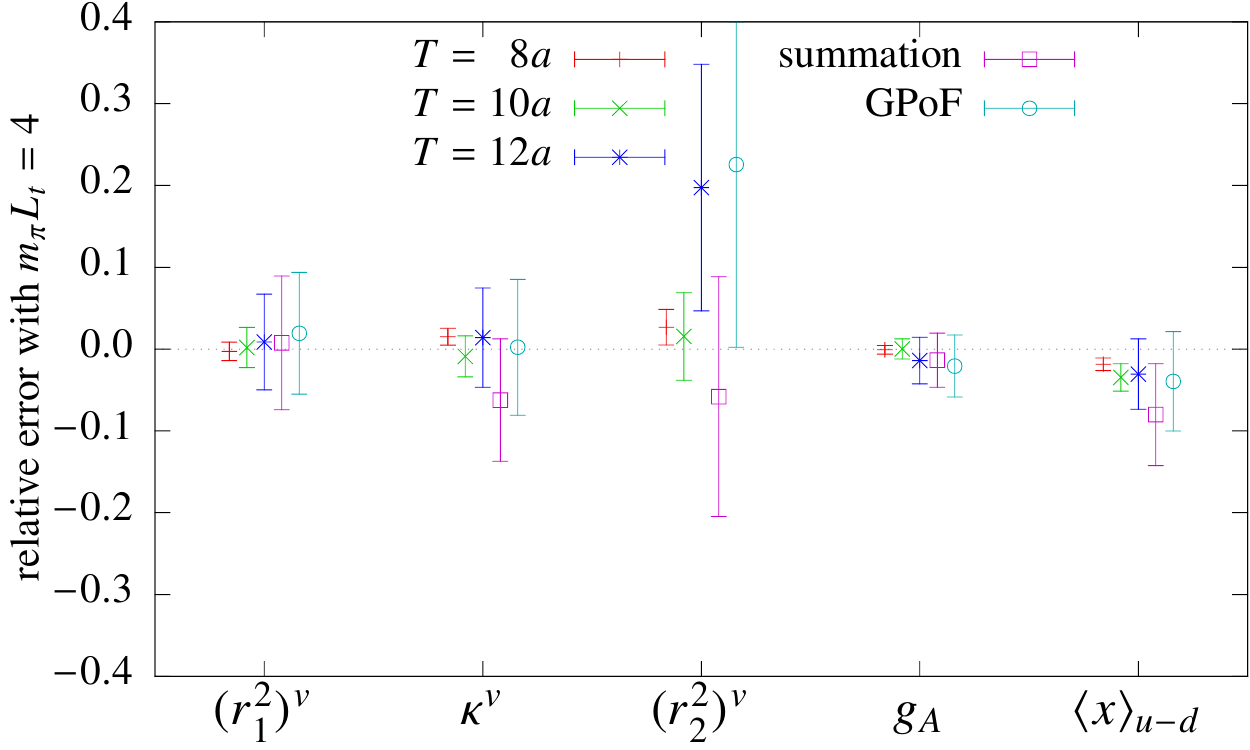}
  \caption{Relative finite-volume errors from ensembles with
    $m_\pi=254$~MeV, interpolated to $m_\pi L_s=4$ and $m_\pi
    L_t=4$. In terms of the fit parameters in
    Eq.~\protect\ref{eq:corners_fit}, the plotted data are $e^{-4}B/A$ and
    $e^{-4}C/A$.}
  \label{fig:corners}
\end{figure}

These results are summarized in Fig.~\ref{fig:corners}, where we also
show the axial charge $g_A$ and the isovector average quark momentum
fraction $\langle x\rangle_{u-d}$. The former shows finite-volume
effects consistent with zero and (at the one-sigma level) smaller than
those observed in Ref.~\cite{Yamazaki:2008py}; using the shortest
source-sink separation suggests that such effects are smaller than 1\%
when $m_\pi L=4$. On the other hand, the latter does show signs of
finite-volume effects, with shifts in opposite directions caused by
finite spatial and temporal extents.

\section{Conclusions}

It has become clear that the use of multiple source-sink separations
for identifying and controlling the presence of excited-state
contamination is essential for a realistic calculation of nucleon
structure observables using lattice QCD. Applying the summation method
on an ensemble with a near-physical pion mass, we calculated isovector
electric and magnetic form factors that are consistent with
experiment.

Using a fully-controlled study of finite-volume effects at
$m_\pi=254$~MeV, we found no such effects for the isovector Dirac
radius or the axial charge, at the presently-available level of
precision, and an indication of some possible non-negligible effects
for the isovector Pauli radius, anomalous magnetic moment, and quark
momentum fraction.

\acknowledgments

We thank Zoltan Fodor for useful discussions and the
Budapest-Marseille-Wuppertal collaboration for making some of their
configurations available to us. This research used resources of the
Argonne Leadership Computing Facility at Argonne National Laboratory,
which is supported by the Office of Science of the U.S.\ Department of
Energy under contract \#DE--AC02--06CH11357, resources at
Forschungszentrum Jülich, and facilities of the USQCD Collaboration,
which are funded by the Office of Science of the U.S.\ Department of
Energy.

During this research JG, SK, SM, JN, AP and SS were supported in part
by the U.S.\ Department of Energy Office of Nuclear Physics under
grant \#DE--FG02--94ER40818, ME was supported in part by DOE grant
\#DE--FG02--96ER40965, SS was supported in part by DOE contract
\#DE--AC02--05CH11231, SK was supported in part by Deutsche
Forschungsgemeinschaft through grant SFB--TRR~55, and JG was supported
in part by the PRISMA Cluster of Excellence at the University of
Mainz.

Calculations were performed with Qlua~\cite{Qlua}, except for the
propagator solves for the high-precision study of excited-state
effects, which were done with the Chroma software
suite~\cite{Edwards:2004sx}, using QUDA~\cite{Clark:2009wm} with
multi-GPU support~\cite{Babich:2011:SLQ:2063384.2063478}.

\bibliographystyle{JHEP-2-notitle}
\bibliography{formfactors.bib}

\providecommand{\href}[2]{#2}\begingroup\raggedright\begin{thebibliography}{10}

\bibitem{Beane:2012vq}
S.~R. Beane, E.~Chang, S.~D. Cohen, W.~Detmold, H.~W. Lin {\em et~al.}, {\em
  Phys. Rev. D} {\bf 87} (2013) 034506
  [\href{http://arXiv.org/abs/1206.5219}{{\tt 1206.5219}}].
%%CITATION = ARXIV:1206.5219;%%

\bibitem{Aubin:2010jc}
C.~Aubin and K.~Orginos, {\em AIP Conf. Proc.} {\bf 1374} (2011) 621--624
  [\href{http://arXiv.org/abs/1010.0202}{{\tt 1010.0202}}].
%%CITATION = ARXIV:1010.0202;%%

\bibitem{Luscher:1990ck}
M.~L\"uscher and U.~Wolff, {\em Nucl. Phys. B} {\bf 339} (1990) 222--252.
%%CITATION = NUPHA,B339,222;%%

\bibitem{Blossier:2009kd}
B.~Blossier, M.~Della~Morte, G.~von Hippel, T.~Mendes and R.~Sommer, {\em JHEP}
  {\bf 0904} (2009) 094 [\href{http://arXiv.org/abs/0902.1265}{{\tt
  0902.1265}}].
%%CITATION = ARXIV:0902.1265;%%

\bibitem{Capitani:2010sg}
S.~Capitani, B.~Knippschild, M.~Della~Morte and H.~Wittig, {\em PoS} {\bf
  LATTICE2010} (2010) 147 [\href{http://arXiv.org/abs/1011.1358}{{\tt
  1011.1358}}].
%%CITATION = ARXIV:1011.1358;%%

\bibitem{Bulava:2010ej}
J.~Bulava, M.~A. Donnellan and R.~Sommer, {\em PoS} {\bf LATTICE2010} (2010)
  303 [\href{http://arXiv.org/abs/1011.4393}{{\tt 1011.4393}}].
%%CITATION = ARXIV:1011.4393;%%

\bibitem{Durr:2010aw}
S.~D\"urr, Z.~Fodor, C.~Hoelbling, S.~Katz, S.~Krieg {\em et~al.}, {\em JHEP}
  {\bf 1108} (2011) 148 [\href{http://arXiv.org/abs/1011.2711}{{\tt
  1011.2711}}].
%%CITATION = ARXIV:1011.2711;%%

\bibitem{Green:2012ud}
J.~R. Green, M.~Engelhardt, S.~Krieg, J.~W. Negele, A.~V. Pochinsky {\em
  et~al.}, \href{http://arXiv.org/abs/1209.1687}{{\tt 1209.1687}}.
%%CITATION = ARXIV:1209.1687;%%

\bibitem{Alberico:2008sz}
W.~M. Alberico, S.~M. Bilenky, C.~Giunti and K.~M. Graczyk, {\em Phys. Rev. C}
  {\bf 79} (2009) 065204 [\href{http://arXiv.org/abs/0812.3539}{{\tt
  0812.3539}}].
%%CITATION = ARXIV:0812.3539;%%

\bibitem{Yamazaki:2008py}
T.~Yamazaki, Y.~Aoki, T.~Blum, H.~W. Lin, M.~F. Lin {\em et~al.}, {\em Phys.
  Rev. Lett.} {\bf 100} (2008) 171602
  [\href{http://arXiv.org/abs/0801.4016}{{\tt 0801.4016}}].
%%CITATION = ARXIV:0801.4016;%%

\bibitem{Qlua}
A.~Pochinsky, ``Qlua.''
  \href{https://usqcd.lns.mit.edu/qlua}{\texttt{https://usqcd.lns.mit.edu/qlua}}.

\bibitem{Edwards:2004sx}
R.~G. Edwards and B.~Jo\'o, {\em Nucl. Phys. Proc. Suppl.} {\bf 140} (2005) 832
  [\href{http://arXiv.org/abs/hep-lat/0409003}{{\tt hep-lat/0409003}}].
%%CITATION = HEP-LAT/0409003;%%

\bibitem{Clark:2009wm}
M.~Clark, R.~Babich, K.~Barros, R.~Brower and C.~Rebbi, {\em Comput. Phys.
  Commun.} {\bf 181} (2010) 1517--1528
  [\href{http://arXiv.org/abs/0911.3191}{{\tt 0911.3191}}].
%%CITATION = ARXIV:0911.3191;%%

\bibitem{Babich:2011:SLQ:2063384.2063478}
R.~Babich, M.~A. Clark, B.~Jo\'{o}, G.~Shi, R.~C. Brower and S.~Gottlieb, in
  {\em Proceedings of 2011 International Conference for High Performance
  Computing, Networking, Storage and Analysis}, SC '11, (New York, NY, USA),
  pp.~70:1--70:11, ACM, 2011.
\newblock \href{http://arXiv.org/abs/1109.2935}{{\tt 1109.2935}}.
%%CITATION = ARXIV:1109.2935;%%

\end{thebibliography}\endgroup

\end{document}